\documentclass[namedreferences]{SolarPhysics}
\usepackage[optionalrh]{spr-sola-addons} 
\usepackage{graphicx}        
\usepackage{color}           
\usepackage{url}             

\usepackage{solaheader}	
\newcommand{\solareceiveddate}{7 July 2010} 
\newcommand{\solaaccepteddate}{31 July 2010}

\newcommand{\adv}{    {\it Adv. Space Res.}}

\newcommand{\apj}{    {\it Astrophys. J.}}

\newcommand{\pasj}{   {\it Pub. Astron. Soc. Japan}}

\newcommand{\solphys}{{\it Solar Phys.}}

\begin{document}

\begin{article}

\begin{opening}

\title{Time-dependent Stochastic Modeling of Solar Active Region 
  Energy}

\author{M.~Kanazir and M.~S.~\surname{Wheatland$^{1}$}}
\runningauthor{M.~Kanazir and M.~S. Wheatland}
\runningtitle{Time-dependent Stochastic Modeling of Active Region 
  Energy}

\institute{$^{1}$ Sydney Institute for Astronomy, School of 
  Physics, The University of Sydney, NSW 2006, Australia \\
email: \url{michael.wheatland@sydney.edu.au}}

\begin{abstract}
A time-dependent model for the energy of a flaring solar 
active region is presented based on an existing 
stochastic jump-transition model \cite{1998ApJ...494..858W,2008ApJ...679.1621W,2009SoPh..255..211W}. 
The magnetic 
free energy of an active region is assumed to vary in time due to a 
prescribed (deterministic) rate of energy input and prescribed (random) 
jumps downwards in energy due to flares. The existing 
model reproduces observed flare statistics, in particular 
flare frequency-size and waiting-time distributions, but modeling 
presented to date has considered only the time-independent choices of 
constant energy input and constant flare transition rates with a
power-law distribution in energy. These choices may be appropriate for 
a solar active region producing a constant mean rate of flares. 
However, many solar active regions exhibit time variation in their 
flare productivity, as exemplified by NOAA active region AR 11029, 
observed during October-November 2009 \cite{2010ApJ...710.1324W}. 
Time variation is incorporated into the jump-transition model for 
two cases: 1.\ a step change in the rates of flare transitions; 
and 2.\ a step change in the rate of energy supply to the 
system. Analytic arguments are presented describing the qualitative 
behavior of the system in the two cases. In each case the system 
adjusts by shifting to a new stationary state over a relaxation time 
which is estimated analytically. The model exhibits 
flare-like event statistics. In each case the frequency-energy 
distribution is a power law for flare energies less than a 
time-dependent rollover set by the largest energy the system is 
likely to attain at a given time. The rollover is not observed if the 
mean free energy of the system is sufficiently large. For Case~1, the 
model exhibits a double exponential waiting-time distribution, 
corresponding to flaring at a constant mean rate during two intervals 
(before and after the step change), if the average energy of the 
system is large. For Case~2 the waiting-time distribution is a 
simple exponential, again provided the average energy of the system 
is large. Monte Carlo simulations of Case~1 are presented which 
confirm the estimate for the relaxation time, and confirm the expected 
forms of the frequency-energy and waiting-time distributions. The 
simulation results provide a qualitative model for observed flare 
statistics in active region AR 11029.
\end{abstract}
\keywords{Active Regions, Models; 
Corona, Models; Flares, Models; Flares, Microflares and Nanoflares}
\end{opening}


\section{Introduction}
     \label{sec:Introduction} 

Solar flares are magnetic explosions in the Sun's outer atmosphere,
the corona, which occur intermittently. Local space weather effects
due to large flares, including damage to electronics on expensive 
communications satellites, motivate attempts to predict flare 
occurrence \cite{2006AdSpR..38..280O,2008sswe.rept.....C}. However, 
existing prediction methods are probabilistic, and not very reliable 
\cite{2005SpWea...307003W,2008ApJ...688L.107B}.

The flare mechanism is believed to be magnetic reconnection, which 
occurs at certain topological sites in the coronal magnetic field
configuration of active regions around sunspots 
\cite{2002A&ARv..10..313P}. However, the reconnection process is 
incompletely understood, and detailed observations of individual 
flares reveal a diversity of effects which are often secondary 
phenomena, providing only indirect information on the underlying 
flare mechanism \cite{2008LRSP....5....1B}. 
The statistics of solar flare occurrence have been 
examined in an attempt to provide insight into the flare phenomenon, 
and also to improve flare prediction.

\subsection{Observed flare statistics}
  \label{subsec:ObsStats}

Two statistical distributions of interest are the flare
frequency-size distribution, and the waiting-time distribution. The 
frequency-size distribution ${\cal N}(S)$ is the number of events 
per unit time and per unit size $S$, where `size' refers to a measure 
of the magnitude of an event, for example the peak flux in X-ray 
wavelengths, or an estimate of the energy. The frequency-size 
distribution is observed to be a featureless power-law over many 
orders of magnitude \cite{1956PASJ....8..173A,2005psci.book.....A}:
\begin{equation}\label{eq:freq_size} 
{\cal N}(S)=\lambda_1 (\gamma -1 )S_1^{\gamma-1} S^{-\gamma},
\end{equation} 
where $\gamma \approx 1.5$--2 is the power-law index (the exact value
depends on the choice of the measure for size)
and $\lambda_1$ is the total mean rate of flaring for events larger
than $S_1$. 
Few exceptions to this simple power-law distribution have been reported 
[see however \inlinecite{2010ApJ...710.1324W}], although an upper 
rollover or cut-off is required on energetics grounds 
\cite{1991SoPh..133..357H}. 
The waiting-time distribution is the distribution $P(\Delta t)$ of times 
$\Delta t$ between flare events. For individual active regions the 
waiting-time distribution is often observed to be consistent with
a simple exponential:
\begin{equation}\label{eq:wtd_poiss}
P(\Delta t)= \lambda_1 {\rm e}^{-\lambda_1\Delta t}.
\end{equation}
where $\lambda_1$ is the mean rate of events of events defined by 
Equation~(\ref{eq:freq_size}). 
This model corresponds to flares occurring as independent 
random events at a constant mean rate, i.e.\ as a Poisson process in 
time \cite{2001JGR...10629951M,2001SoPh..203...87W}. However, the mean
rate of flaring in active regions is often observed to vary, in which
case a piecewise-constant Poisson model may be appropriate. The 
model waiting-time distribution is then a sum of exponentials 
corresponding to distinct intervals with different rates 
\cite{2002SoPh..211..255W}:
\begin{equation}\label{eq:wtd_pconst-poiss}
P(\Delta t)= \sum_i\frac{n_{1i}}{N_1}\lambda_{1i} 
  {\rm e}^{-\lambda_{1i}\Delta t},
\end{equation}
where $n_{1i}=\lambda_{1i} t_i$ is the number of events with size 
larger than $S_1$ corresponding to a rate $\lambda_{1i}$ and an
interval $t_i$, and where $N_1=\sum_i n_{1i}$ is the total number of 
events. A power-law tail is observed in 
the combined waiting-time distribution for events from multiple active 
regions over longer periods of time on the Sun 
\cite{1999PhRvL..83.4662B}, which also may be accounted for by the 
time-dependent Poisson model 
\cite{2000ApJ...536L.109W,2002SoPh..211..255W}, although some authors 
have argued for a non-Poisson interpretation \cite{2001ApJ...555L.133L}.
Recently \inlinecite{2010ApJ...717..683A} demonstrated that a variety
of observations for the longer-term waiting-time distribution are
consistent with Poisson occurrence in time according to a simple
functional form for the distribution of rates in time, which is a 
variant of the rate distribution presented in 
\inlinecite{2000ApJ...536L.109W}.

\subsection{Flare statistics in two active regions}
  \label{subsec:TwoARs}

To illustrate these ideas we consider two active regions: US National
Oceanic and Atmospheric Administration (NOAA) active region AR 10486, 
from October-November 2003, and NOAA AR 11029, from October-November 
2009. The data used are the soft X-ray event lists compiled by the US 
Space Weather Prediction Center 
(SWPC)\footnote{See http://www.swpc.noaa.gov/.}. The events are 
selected from whole-Sun $1$--$8\,\mbox{\AA}$ flux measurements by the 
Geostationary Observational Environmental satellites (GOES). The peak 
flux of GOES events is routinely used to classify flares, and is the 
measure of size used here. The SWPC/GOES data are often used in studies 
of flare statistics, although they are less than ideal for this 
purpose because the lists are incomplete, and the peak fluxes in the 
lists are not background-subtracted [for a discussion see e.g.\ 
\inlinecite{2001SoPh..203...87W}]. However, the events for active
region AR 
11029 shown here are individually background subtracted, based on work 
in an earlier study~\cite{2010ApJ...710.1324W}.

\begin{figure}[here]
\centerline{\includegraphics[width=1\textwidth]{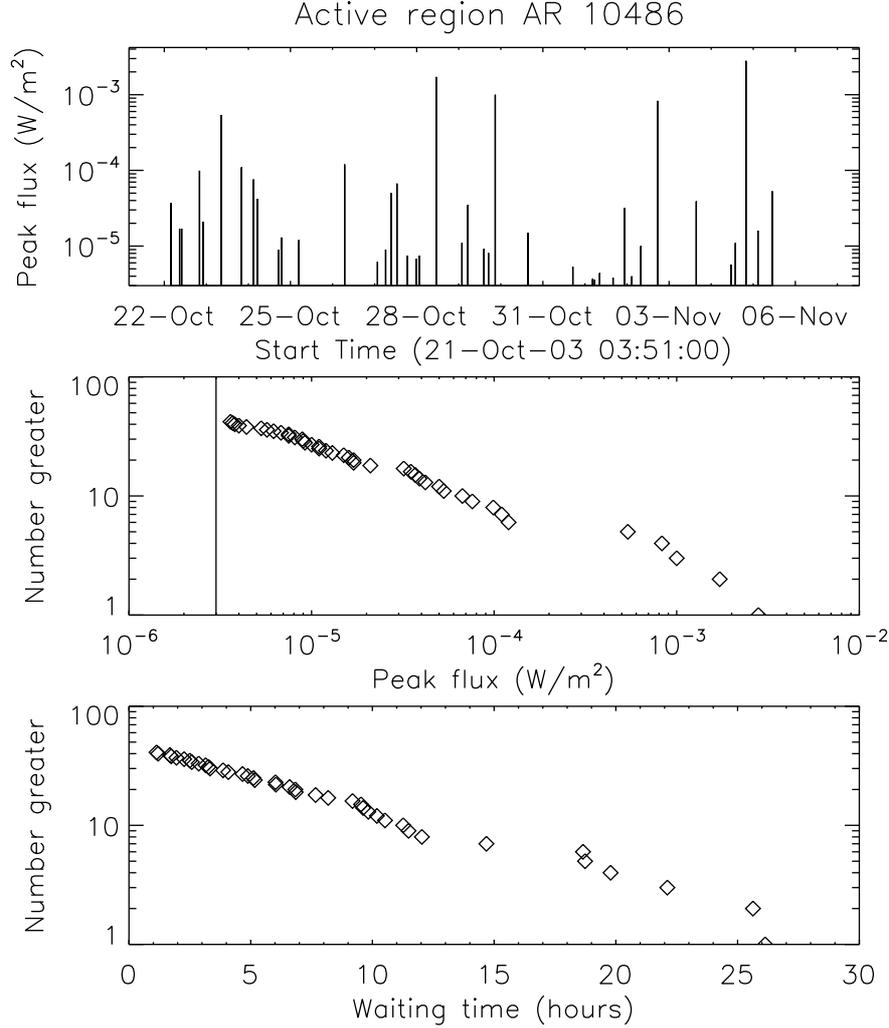}}
\caption{Flares in the GOES events lists for NOAA active region 
AR 10486 above peak soft X-ray flux 
$3\times 10^{-6}{\rm W}{\rm m}^{-2}$ (43 events).
Upper panel: Schematic showing the peak flux of each event at the 
event time. Middle panel: The frequency-peak flux distribution, as a 
cumulative plot in a log-log representation. Lower panel: The 
waiting-time distribution as a cumulative plot in a log-linear
representation.}
\label{fig:dists_ar10486}
\end{figure}

Figure~\ref{fig:dists_ar10486} illustrates the frequency-energy, and
waiting-time distributions, for soft X-ray flares observed in
active region AR 10486. This large and highly complex sunspot region
appeared in the late stages of the maximum of the last solar cycle 
and produced a remarkable sequence of extremely large solar flares
\cite{2007ApJ...657..577D,2008ARep...52..852C}.
Figure~\ref{fig:dists_ar10486} shows the events in this region larger 
than peak flux $S_1=3\times 10^{-6}\,{\rm W}\,{\rm m}^{-2}$ (this 
choice should ensure the list of events is relatively complete, for 
events larger than $S_1$). The region produced $N_1=43$ events with
peak flux larger than $S_1$, including the biggest flare of the modern 
era (with listed peak flux
$2.8\times 10^{-3}\,{\rm W}\,{\rm m}^{-2}$), on 4 November 2003.  
The upper panel shows the peak fluxes versus the recorded peak times 
for the events as a sequence of vertical lines, with a logarithmic 
scaling for flux, the 4 November flare being the tallest line. The 
middle panel plots the the number of events above size $S$ versus size, 
which corresponds to the cumulative peak-flux distribution $C(S)$, 
defined in terms of the frequency-peak flux distribution by 
\begin{equation}\label{eq:cum_size}
C(S)=\frac{T}{N_1}\int_S^{\infty}{\cal N}(S^{\prime})\,dS^{\prime}, 
\end{equation}
where $T$ is the duration of the observing interval, and $S>S_1$. 
The vertical line in this panel indicates $S_1$. 
The events are approximately power-law distributed in peak flux (a 
straight line in this representation), consistent with 
Equation~(\ref{eq:freq_size}). The lower panel plots the waiting-time
distribution for the same events, again as a cumulative distribution
i.e.\ the number of waiting times larger than a given time versus 
waiting time. This corresponds to the integral of $P(\Delta t)$ with
respect to $\Delta t$. The distribution is presented with a 
log-linear scaling, and reveals that the events are approximately 
exponentially distributed (a straight line, in this representation),
consistent with Equation~(\ref{eq:wtd_poiss}). 

Figure~\ref{fig:dists_ar10486} shows 
that AR 10486, although remarkable in terms of the size of particular 
events, was unremarkable statistically. It produced flares according 
to a power-law frequency-peak flux distribution at a constant mean 
rate during its transit of the disk. This suggests that the physical
conditions underlying flaring are approximately time-independent.

\begin{figure}[here]
\centerline{\includegraphics[width=1\textwidth]{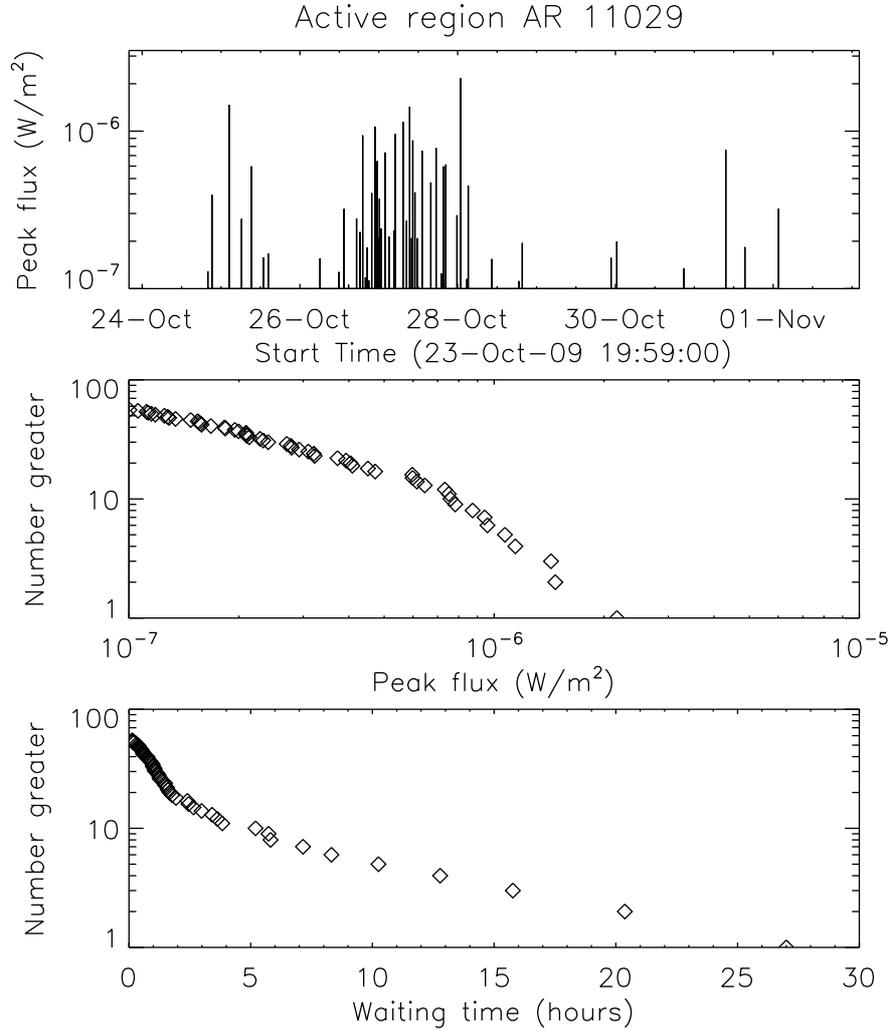}}
\caption{Flares in the GOES events lists for NOAA active region 
AR 11029 with a background-subtracted peak flux above
$10^{-7}\,{\rm W}{\rm m}^{-2}$, a total of 56 events 
(Wheatland 2010). The presentation of the data is the same 
as in Figure~1.}
\label{fig:dists_ar11029}
\end{figure}

Figure~\ref{fig:dists_ar11029} shows GOES event data for active
region AR 11029, which exhibits time variation in its mean rate of 
flaring. This spatially small but highly flare-productive active region
emerged on the disk in October 2009, during an extended interval of 
low solar activity. The statistics of events in this region were 
investigated in
\inlinecite{2010ApJ...710.1324W}. The presentation of 
Figure~\ref{fig:dists_ar11029} is the same as 
Figure~\ref{fig:dists_ar10486}, but in this case the data are
individually background-subtracted, which is particularly important 
because the events are small. There are $N_1=56$ events above a
backgrounded-subtracted peak flux of 
$S_1=10^{-7}\,{\rm W}\,{\rm m}^{-2}$, as shown in the upper panel.
This panel suggests that the flaring rate is high for two days (26 
October and 27 October), and relatively low at other times, an 
interpretation supported by a Bayesian rate analysis 
\cite{1998ApJ...504..405S,2010ApJ...710.1324W}. The changes in the rate
are quite dramatic: a relatively sudden increase by a factor of about
ten, and a sudden decrease by about the same factor. 
The lower panel plots the cumulative waiting-time distribution, 
which shows a double exponential form, consistent with Poisson 
occurrence at two different mean rates (a low and a high rate), as 
represented by Equation~(\ref{eq:wtd_pconst-poiss}) with two 
intervals/rates. The middle panel in 
Figure~\ref{fig:dists_ar11029} plots the cumulative peak-flux 
distribution for the flares in AR 11029, and is suggestive of a 
rollover around $10^{-6}\,{\rm W}\,{\rm m}^{-2}$. 
\inlinecite{2010ApJ...710.1324W} applied Bayesian model comparison to
show that a power-law plus upper rollover model is much more probable
for this data set than a simple power-law model, and argued that the 
departure from a power law may reflect the finite storage of magnetic 
energy in this (small) region. 

Time variation in mean flaring rate is commonly observed in active 
regions \cite{2001SoPh..203...87W}, and is often associated, as in the 
case of active region AR 11029, with a change in the photospheric 
magnetic complexity of a region \cite{2010ApJ...710.1324W}. On October
26 the region increased in photospheric magnetic complexity 
(becoming a $\beta$--$\gamma$ region, in the Mt Wilson classification), 
coincident with the increase in flaring rate. 
Photospheric magnetic field changes are likely to be reflected in 
changes in the magnetic field configuration in the corona, which may 
facilitate or inhibit reconnection, and hence change 
the flaring rate. It is plausible that the more interesting statistics 
observed for AR 11029, by comparison with AR 10486 
(Figure~\ref{fig:dists_ar10486}) are related to the smaller
physical smaller size of the region, which may make the coronal 
magnetic field configuration more sensitive to change.

\subsection{Models for flare statistics}
  \label{subsec:ModStats}

Models for solar flare statistics attempt to account for the observed
distributions. Most of the models are motivated either by accounting 
for energy balance in an active region 
\cite{1978ApJ...222.1104R,1994SoPh..151..195L,2001SoPh..202..109C} 
or by explaining the power law in the frequency-size distribution 
using self-organized criticality/cellular automata (`avalanche' 
models) 
\cite{1991ApJ...380L..89L,2001SoPh..203..321C,2003PhRvL..90m1101H}, 
although the two pictures are not mutually exclusive. The avalanche 
model produces a power-law frequency-size distribution below an
upper rollover set by the size of the cellular automata 
\cite{1993ApJ...412..841L}, and has an exponential (Poisson) 
waiting-time distribution, if the system is subject to a constant 
mean rate of driving \cite{1994PhDT........51B,1998ApJ...509..448W}, 
but may produce other waiting-time distributions with time-dependent 
driving \cite{2001ApJ...557..891N}. Energy balance models are 
designed to explain the power-law frequency-energy distribution, and 
generally assume Poisson flare occurrence, but simple models in which
active region energy is completely removed by a flare imply
a relationship between waiting times and energy, which is not observed
\cite{1995ApJ...447..416L}. 
A generalized energy-balance formalism (a `jump-transition model') 
which avoids this problem was 
introduced by \inlinecite{1998ApJ...494..858W}, and further developed
in \inlinecite{2008ApJ...679.1621W} and 
\inlinecite{2009SoPh..255..211W}. The jump-transition model is applied 
in this paper.
 
The general jump-transition model 
\cite{1998ApJ...494..858W,2008ApJ...679.1621W,2009SoPh..255..211W} 
describes the magnetic free energy $E=E(t)$ of an active region at 
time $t$ in terms of secular energy input at a rate $\beta (E,t)$, and
stochastic transitions downwards in energy (flares) at a rate 
$\alpha (E,E^{\prime},t)$ per unit time and per unit energy for
transitions from $E$ to $E^{\prime}$. The model may be presented
either in terms of an integro-partial differential equation describing 
the probability distribution $P(E,t)$ for the energy 
\cite{1998ApJ...494..858W,2008ApJ...679.1621W}, called a
`master equation' \cite{1992sppc.book.....V,Gardiner2004}, or in terms 
of a stochastic differential equation describing the time evolution of
$E=E(t)$ \cite{2009SoPh..255..211W}. 

Modeling using the jump-transition formalism to date has considered 
only time-independent cases, namely a time- and energy-independent 
rate of energy input:
\begin{equation}\label{eq:beta_wg98}
\beta (E) = \beta_0, 
\end{equation}
where $\beta_0$ is a
constant, and time-independent power-law distributed flare 
transitions\footnote{\inlinecite{2008ApJ...679.1621W} and 
\inlinecite{2009SoPh..255..211W} considered also the
case with an additional factor $E^{\delta}$ multiplying the transition
rates given by Equation~(\ref{eq:pl_alpha_wg98}), but the results 
suggested that Equation~(\ref{eq:pl_alpha_wg98}) is the preferred 
model.}:
\begin{equation}\label{eq:pl_alpha_wg98}
\alpha (E,E^{\prime})=\alpha_0 (E-E^{\prime})^{-\gamma} \theta
(E-E^{\prime}-E_c),
\end{equation} 
where $\alpha_0$ is a constant, $E_c$ is a low-energy 
cutoff, $\theta (x)$ denotes the step 
function, and $\gamma =1.5$ is the observed power-law index in the 
flare frequency-energy distribution \cite{2005psci.book.....A}.
The model flare frequency-energy distribution is given by
\begin{eqnarray}\label{eq:ffe_pl_alpha}
{\cal N}(E)&=&\int_{E}^{\infty}P(E^{\prime})
  \alpha (E^{\prime},E^{\prime}-E ){\mathrm d}E^{\prime}
  \nonumber \\
  &=&\alpha_0 E^{-\gamma}
  \int_{E}^{\infty}P(E^{\prime}){\mathrm d}E^{\prime}
  \quad \mbox{for} \quad E\geq E_c,
\end{eqnarray}
and is zero for $E<E_c$. Equation~(\ref{eq:ffe_pl_alpha}) describes a 
power law with index $\gamma$ above $E_c$ and below
an upper rollover set by the largest energy the system is likely to
attain, which has an approximate lower bound given by the estimate 
for the average or mean energy of the system \cite{2008ApJ...679.1621W}:
\begin{equation}\label{eq:Ea}
{\cal E} = \left(\frac{2-\gamma}{\alpha_0/\beta_0}
  \right)^{1/(2-\gamma)}.
\end{equation}
For a sufficiently large rate of energy supply with respect to the 
rate of flare transitions, we have
\begin{equation}\label{eq:largeEcriterion}
\frac{\beta_0}{\alpha_0 E_c^{2-\gamma}}\gg 1,
\end{equation}
and hence ${\cal E}\gg E_c$, i.e.\ the system mean energy is large.
In that case most flares do not significantly deplete the free energy,
and then, according to Equation~(\ref{eq:ffe_pl_alpha}), the 
observed flare frequency-energy distribution is a simple power law.
The rollover is not relevant unless a very long time history of flaring
(including infrequent, very large events) is observed.
In the case ${\cal E}\gg E_c$ also, flares occur
as a Poisson process in time, i.e.\ the waiting-time distribution is
a simple exponential. This may be understood by noting that the total 
flaring rate assuming the system has energy $E$ is 
\begin{eqnarray}\label{eq:lambda_tind}
\lambda (E)&=&
  \int_0^{E}\alpha (E,E^{\prime}){\mathrm d}E^{\prime}
  \nonumber \\
  &=& \alpha_0
    \left(E_c^{-\gamma+1}-E^{-\gamma+1}\right)/(\gamma-1) \quad
      \mbox{for} \quad E\geq E_c,
\end{eqnarray}
and $\lambda (E) = 0$ for $E<E_c$.
When ${\cal E}\gg E_c$ we can assume $E\gg E_c$, in which case 
Equation~(\ref{eq:lambda_tind}) is well approximated by
\begin{equation}\label{eq:lambda_W08_largeE}
\lambda (E) =
  \frac{\alpha_0}{\gamma-1} E_c^{-\gamma+1}.
\end{equation}
The right-hand side of Equation~(\ref{eq:lambda_W08_largeE}) is 
energy- and hence time-independent, so flares occur as a 
simple Poisson process in time. However, if flares deplete the energy 
of the system, then the system energy may become comparable to $E_c$, 
in which case the total mean flaring rate depends on the energy of 
the system and varies in time, and the waiting-time distribution 
departs from a simple exponential 
\cite{2008ApJ...679.1621W,2009SoPh..255..211W}. These comments show 
how observed flare statistics provide insight into magnetic energy 
balance in an active region, in the context of a model. 

The jump-transition model is a general formalism describing 
time-dependent as well as stationary situations. In this paper we 
apply the model to time-dependent cases for the first time, to
investigate how time variation affects the model event 
statistics, in particular the frequency-energy and waiting-time 
distributions. We focus on two cases: a sudden change in the 
flaring rate; and a sudden change in the energy supply rate. The
first case may be appropriate to describe, e.g., active 
region AR 11029 on 26 October 2009 (see the upper panel of 
Figure~\ref{fig:dists_ar11029}). The second case
may be appropriate to describe an active region which starts to grow
as a result of the emergence of new magnetic flux, but which retains 
a given magnetic configuration and does not change its flaring
rate.

The layout of
the paper is as follows. Section~\ref{sec:Model} briefly re-iterates
the details of the jump-transition formalism 
(Section~\ref{subsec:MEandSDE}), describes the flare-like choices
for modeling time-variation considered here 
(Section~\ref{subsec:TDepChoices}), and then presents analytic
arguments which allow the general behavior of the system to be deduced
(Section~\ref{subsec:AnalyticTDep}). 
Section~\ref{sec:Stochastic} presents Monte Carlo simulations of the 
time-dependent system for the case of a sudden increase in the flaring
rate, which provide a qualitative model for the observed behavior of 
active region AR 11029. A brief description of the numerical methods 
is given (Section~\ref{subsec:NumMeth}), followed by an account of 
the results (Section~\ref{subsec:Results}). 
Section~\ref{sec:Conclusions} presents conclusions.

\section{Model}
     \label{sec:Model} 
    
\subsection{General master equation and stochastic differential 
equation models}
 \label{subsec:MEandSDE} 
 
\inlinecite{1998ApJ...494..858W}, \inlinecite{2008ApJ...679.1621W},
and \inlinecite{2009SoPh..255..211W} developed a stochastic 
jump-transition model for the free magnetic energy of a solar active 
region, which is described by a master equation, or an equivalent 
stochastic differential equation~\cite{1992sppc.book.....V,Gardiner2004}.

In the master equation approach, the probability distribution $P(E,t)$
for the energy $E$ of the system at time $t$ is given by the solution
of the integro-partial differential equation 
\begin{eqnarray}\label{eq:master}
\frac{\partial P(E,t)}{\partial t} &=&
-\frac{\partial }{\partial E}\left[\beta (E,t) P(E,t)\right]
-\lambda (E,t)P(E,t)
\nonumber \\ &+&\int_E^{\infty}P(E^{\prime},t)\alpha
(E^{\prime},E,t){\mathrm d}E^{\prime},
\end{eqnarray}
where $\beta (E,t)$ is the rate of energy input to the system, 
$\alpha (E,E^{\prime},t)$ is the rate of flare jumps in energy from
$E$ to $E^{\prime}$ per unit energy, and
\begin{equation}\label{eq:lambda}
\lambda (E,t)=\int_0^{E}\alpha
(E,E^{\prime},t){\mathrm d}E^{\prime}
\end{equation}
is the total rate of flaring when the system has energy $E$. If the
energy supply rate and the transition rates do not vary with time then
the time-independent version of Equation~(\ref{eq:master}) applies 
(obtained by setting $\partial/\partial t =0$).

The observable distributions of interest are the frequency-energy
distribution and the waiting-time distribution. The model 
frequency-energy distribution is
\begin{equation}\label{eq:ffe}
{\cal N}(E,t)=\int_{E}^{\infty}P(E^{\prime},t)\alpha
(E^{\prime},E^{\prime}-E,t){\mathrm d}E^{\prime}.
\end{equation}
In the time-independent case, the model waiting-time distribution may
be expressed in terms of the stationary probability 
distribution $P(E)$ for the system energy and the solution to an 
auxiliary first-order partial differential equation, as shown by
\inlinecite{2007PhRvE..75a1119D} in the context of general 
jump-transition modeling. The details of this procedure are given in 
\inlinecite{2008ApJ...679.1621W} and \inlinecite{2009SoPh..255..211W} 
but are omitted here. In the general time-dependent case there does not 
appear to be a straightforward way to obtain the waiting-time
distribution from the master equation approach.

\inlinecite{1998ApJ...494..858W} and 
\inlinecite{2008ApJ...679.1621W} solved
Equation~(\ref{eq:master}) for the time-independent 
flare-like choices discussed in Section~\ref{sec:Introduction}, namely
a constant rate of energy supply and power-law distributed transition 
rates [Equations~(\ref{eq:beta_wg98}) and~(\ref{eq:pl_alpha_wg98})],
and the frequency-energy and waiting-time distributions were 
investigated for these choices. An efficient numerical method 
for solving the time-independent master equation was given in 
\inlinecite{2008ApJ...679.1621W}.

Moments of the master equation, i.e.\ averages over energy, provide 
insight into the general behavior of the system 
\cite{2001ApJ...557..332W}. The first moment, obtained by multiplying
Equation~(\ref{eq:master}) by $E$ and integrating with respect to $E$,
gives
\begin{equation}\label{eq:master_moment1}
\frac{d}{dt}\langle E\rangle =\langle\beta\rangle
  -\langle r\rangle , 
\end{equation}
where
\begin{equation}\label{eq:mean_energy}
\langle E\rangle =\int^{\infty}_{0}EP(E,t)dE
\end{equation}
is the mean energy,
\begin{equation}
\langle\beta\rangle =\int^{\infty}_{0}\beta(E,t)P(E,t)dE
\end{equation}
is the mean rate of energy supply, and
\begin{equation}
\langle r\rangle =\int^{\infty}_{0}r(E,t)P(E,t)dE\label{luzer}
\end{equation}
is the mean total rate of loss of energy to flaring, where
\begin{equation}\label{eq:mean_total_loss}
r(E,t)=\int^{E}_{0}(E-E^{\prime})\,\alpha (E,E^{\prime},t)dE^{\prime}
\end{equation}
is the rate of loss of energy due to all jumps. 
Equation~(\ref{eq:master_moment1}) describes how the mean
energy of the system changes in response to time variation in the rate
of energy supply or in the rates of flare transitions.

The equivalent stochastic differential equation to 
Equation~(\ref{eq:master}) is \cite{2007PhRvE..75a1119D}
\begin{equation}\label{eq:stoch_de}
\frac{dE}{dt}=\beta (E,t)-\Lambda (E,t),
\end{equation}
where 
\begin{equation}\label{eq:Lambda}
\Lambda (E,t)=\sum_{i=1}^{N(t)}\Delta E_i\delta (t-t_i)
\end{equation}
is the total loss in energy due to flaring up to time $t$, with 
$\delta (t)$ being the delta function, $N(t)$ being the number of 
events which have occurred up to time $t$, and where the event times 
$t_i$ are 
defined by the Poisson process with time-dependent rate 
$\lambda =\lambda \left[E(t),t\right]$. The factors $\Delta E_i$ are 
the jumps downwards in energy at each flare, which follow the 
distribution $h (\Delta E,E,t)$, defined by
\begin{equation}\label{eq:h}
\alpha (E,E-\Delta E,t )=\lambda (E,t) h (\Delta E,E,t),
\end{equation}
and satisfying the normalization condition
\begin{equation}\label{eq:hnorm}
\int_0^E h(\Delta E,E,t)\,d(\Delta E)=1.
\end{equation}
The stochastic differential equation~(\ref{eq:stoch_de}) is simulated
for the time-independent flare-like choices, using an efficient 
Monte Carlo method, in \inlinecite{2009SoPh..255..211W}.

\subsection{Choices for time variation}
      \label{subsec:TDepChoices} 
      
As a simple time-dependent generalization of the flare-like model
defined by Equations~(\ref{eq:beta_wg98}) and~(\ref{eq:pl_alpha_wg98}),
we consider the choices
\begin{equation}\label{eq:pl_alpha_tdep}
\alpha (E,E^{\prime},t)=\alpha_0(t) (E-E^{\prime})^{-\gamma} \theta
(E-E^{\prime}-E_c)
\end{equation}
with a power-law index $\gamma =1.5$, and
\begin{equation}\label{eq:beta_tdep}
\beta (E,t)=\beta_0 (t).
\end{equation}
Hence we consider time-modulated
flare transition rates with the same power-law functional form for the
change in energy previously considered, and a strictly time-dependent 
energy supply rate. We further restrict attention to the following
two cases.
\begin{itemize}
\item[Case 1:] a step change in the transition-rate coefficient 
$\alpha_0 (t)$,
\begin{equation}\label{eq:step_alpha}
\alpha_0 (t)=\alpha_{01}+(\alpha_{02}-\alpha_{01})\theta (t-T),
\end{equation}
with no change in the energy-supply rate coefficient 
$\beta_0 (t)=\beta_{01}$.
\item[Case 2:] a step change in the energy-supply rate 
coefficient,
\begin{equation}\label{eq:step_beta}
\beta_0 (t)=\beta_{01}+(\beta_{02}-\beta_{01})\theta (t-T),
\end{equation}
with no change in the flare transition-rate coefficient 
$\alpha_0 (t)=\alpha_{01}$.
\end{itemize}
In Equations~(\ref{eq:step_alpha}) and~(\ref{eq:step_beta}) the
factors $\alpha_{0i}$ and $\beta_{0i}$ (with $i=1,2$) are constants,
and $t=T$ denotes the time of the step change.

\begin{figure}[here]
\centerline{\includegraphics[width=1\textwidth]{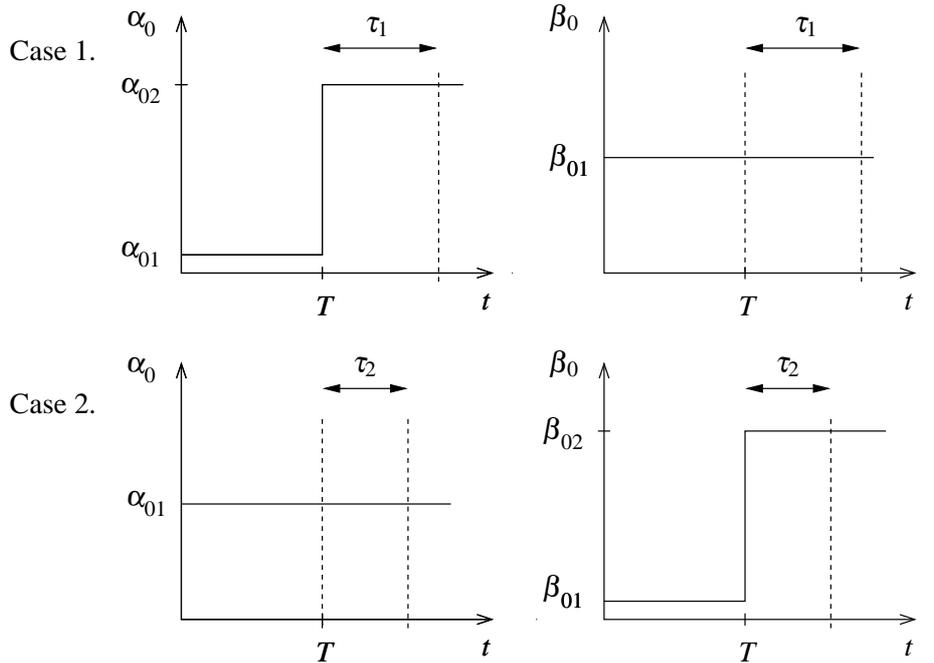}}
\caption{The two simple cases of time variation in the flare model
considered. The upper row is Case 1, a step variation in the flare
transition-rate coefficient $\alpha_0(t)$, and no change in the 
energy supply rate $\beta_0(t)$,
and the lower row is Case 2, a step change in the energy-supply rate
coefficient, and no change in the flare transition rate coefficient. 
The parameters $\tau_1$ and $\tau_2$ are the relaxation 
times, described in Section~\ref{subsec:AnalyticTDep}.}
\label{fig:two_cases}
\end{figure}

Figure~\ref{fig:two_cases} illustrates the two cases, with the upper
row showing Case~1, and the lower row Case~2.
Case 1 may correspond physically to a change in the coronal 
magnetic configuration, which permits enhanced reconnection rates 
to occur, and Case 2 may correspond to increased sub-photospheric 
driving, e.g.\ the emergence of new magnetic flux. 
Case 1 provides a simple model which
may account for the observed behavior of active region AR 11029
on October 26 (see Section~\ref{subsec:TwoARs}, and 
Figure~\ref{fig:dists_ar11029}).

\subsection{Analytic considerations}
 \label{subsec:AnalyticTDep} 
 
For both Case~1 and Case~2 we assume that the system is in a 
steady state prior to
the time $t=T$ of the step change. In other words we assume the system
has had constant flare transition and energy supply 
rates (defined by the coefficients $\alpha_{01}$ and 
$\beta_{01}$ respectively) for an extended interval prior to $t=T$. 
The energy distribution $P_1(E)$ 
describing the steady state is the solution to the time-independent 
master equation with these parameters. The average energy over this 
distribution is given approximately by Equation~(\ref{eq:Ea}), namely
\begin{equation}\label{eq:Emean1}
{\cal E}_{1} = \left(\frac{2-\gamma}{\alpha_{01}/\beta_{01}}
\right)^{1/(2-\gamma )}.
\end{equation}

\begin{figure}[here]
\centerline{\includegraphics[width=0.8\textwidth]{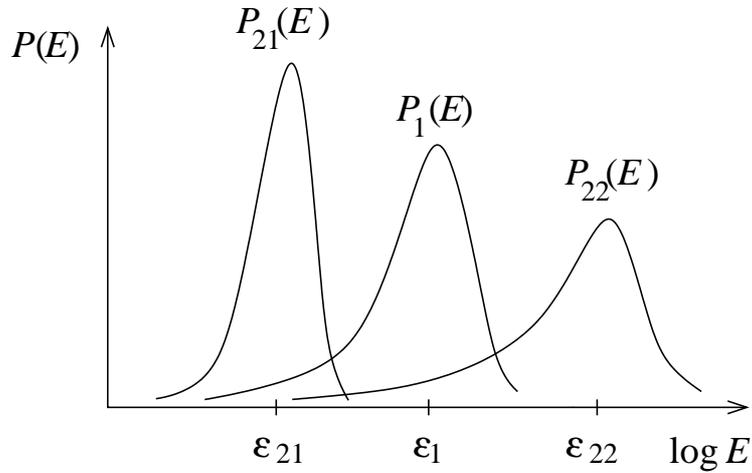}}
\caption{A schematic showing the steady state distribution $P_1(E)$ 
before the step change in parameters, and the steady state 
distributions after relaxation: $P_{21}(E)$ (for Case~1), and 
$P_{12}(E)$ (Case~2). These specific examples correspond 
to Figure~\ref{fig:two_cases}, i.e.\ an increase in the flare
transition rates for Case~1, and an increase in the energy supply 
rate for Case~2.}
\label{fig:dist_shift}
\end{figure}

In both cases the system is in a non-steady state immediately after
the change. In the context of the master equation description, this
means that the energy distribution is inconsistent with the solution
to the time-independent master equation for the new (constant) rate
coefficients. In the context of the stochastic differential equation
description, it means that the energy is not a typical energy for the
system. The evolution of the system in the non-steady state is 
described by a time-dependent energy distribution $P(E,t)$ which is
a solution to the time-dependent master equation~(\ref{eq:master}), 
or by a specific energy trajectory $E=E(t)$ obtained by solving the 
stochastic differential equation~(\ref{eq:stoch_de}). 
A new steady state is
eventually achieved in each case, characterized by new distributions 
$P_{21}(E)$ (for Case~1) and $P_{22}(E)$ (Case~2), which are
solutions to the time-independent master equation with parameters 
defined by rate coefficients $\alpha_{02}$, $\beta_{01}$, and 
$\alpha_{01}$, $\beta_{02}$, respectively. The mean energies of 
these distributions are given approximately by
\begin{equation}\label{eq:Emean21Emean22}
{\cal E}_{21} = \left(\frac{2-\gamma}{\alpha_{02}/\beta_{01}}
\right)^{1/(2-\gamma )} 
\quad \mbox{and} \quad
{\cal E}_{22} = \left(\frac{2-\gamma}{\alpha_{01}/\beta_{02}}
\right)^{1/(2-\gamma )} 
\end{equation}
respectively. The characteristic times for achieving a steady state 
 are referred to as `relaxation' times, and are labeled $\tau_1$ 
and $\tau_2$ for Cases 1 and 2 respectively. These times are
indicated schematically in Figure~\ref{fig:two_cases} by the vertical 
dashed lines. The interval $T<t<T+\tau_i$ (with $i=1,2$) is the
relaxation interval in each case, during which time the mean or peak
of the energy distribution $P(E,t)$ shifts to a new value. 
For the specific examples shown in Figure~\ref{fig:two_cases}, i.e.\ an 
increase in the flare transition rates for Case~1, and an increase 
in the energy-supply rate for Case~2, the peak of the energy 
distribution shifts to a lower energy, and to a higher energy, 
respectively. The locations of the peaks are defined approximately by 
Equations~(\ref{eq:Emean1}) and~(\ref{eq:Emean21Emean22}). 
Figure~\ref{fig:dist_shift} illustrates these changes,
showing the initial steady-state distribution $P_1(E)$ and the final 
steady-state distributions $P_{21}(E)$, and $P_{22}(E)$, and the 
values ${\cal E}_1$, ${\cal E}_{21}$, and ${\cal E}_{22}$.
This is a schematic diagram, not the result of a calculation, but the
distributions have been drawn to approximately match the functional 
forms observed for numerical solutions of the steady-state master equation
\cite{1998ApJ...494..858W,2008ApJ...679.1621W}.

The relaxation times for the two cases may be 
estimated based on the first moment of the master equation, presented
in Section~\ref{subsec:MEandSDE}. Substituting the time-dependent
flare-like choices Equations~(\ref{eq:pl_alpha_tdep}) 
and~(\ref{eq:beta_tdep}) into 
Equations~(\ref{eq:master_moment1})--(\ref{eq:mean_total_loss}) 
gives the more specific form for the moment equation
\begin{equation}\label{eq:master_moment1_flare}
\frac{d}{dt}\langle E\rangle = \beta_0 (t)
- 2\alpha_0(t)\langle E\rangle^{\gamma-1},
\end{equation}
where we assume $E\gg E_c$ and 
$\langle E^{2-\gamma}\rangle \approx \langle E\rangle^{2-\gamma}$.
Equation~(\ref{eq:master_moment1_flare}) is a nonlinear ordinary
differential equation. Implicit analytic forms for solutions may 
be written down for the specific cases of 
interest ($\gamma = 1.5$ and time-independent rate coefficients), 
but it is simplest to estimate a relaxation time directly from 
Equation~(\ref{eq:master_moment1_flare}).

For Case~1, the system shifts from the approximate average energy 
${\cal E}_1$ to the approximate average energy ${\cal E}_{21}$ in 
the relaxation time $\tau_1$ with parameters $\alpha_0(t)=\alpha_{02}$ 
and $\beta_0(t) =\beta_{01}$ during the interval of evolution, so 
Equation~(\ref{eq:master_moment1_flare}) implies the approximate 
relationship
\begin{equation}\label{eq:tau1_working}
\frac{|{\cal E}_{21}-{\cal E}_1|}{\tau_1 }\approx 
\left|\beta_{01} - 2\alpha_{02}{\cal E}_1^{\gamma -1} \right|.
\end{equation}
Substituting Equations~(\ref{eq:Emean1}) and~(\ref{eq:Emean21Emean22})
into Equation~(\ref{eq:tau1_working}) leads to
\begin{equation}\label{eq:tau1}
\tau_1 \approx \frac{\beta_0 (\alpha_{01}+\alpha_{02})}
  {4\alpha_{01}\alpha_{02}^2},
\end{equation}
assuming $\alpha_{02}\neq \alpha_{01}$ and using $\gamma=1.5$. 

Similarly, for Case~2, the system shifts from the approximate average
energy ${\cal E}_1$ to the approximate average energy ${\cal E}_{22}$ in 
time $\tau_2$ with parameters $\alpha_0(t)=\alpha_{01}$ 
and $\beta_0(t) =\beta_{02}$, so Equation~(\ref{eq:master_moment1_flare})
implies
\begin{equation}
\frac{|{\cal E}_{22}-{\cal E}_1|}{\tau_2 }\approx 
\left|\beta_{02} - 2\alpha_{01}{\cal E}_1^{\gamma -1} \right|,
\end{equation}
which leads to
\begin{equation}\label{eq:tau2}
\tau_2 \approx \frac{\beta_{01}+\beta_{02}}{4\alpha_{01}^2},
\end{equation}
assuming $\beta_{02}\neq \beta_{01}$ and using $\gamma=1.5$.
Equations~(\ref{eq:tau1}) and~(\ref{eq:tau2}) are accurate if the
changes are small. If the mean energy of the system increases 
substantially in each case the expressions give overestimates because
of the use of the initial small mean energy to replace the
time-varying average energy $\langle E\rangle$ on 
the right-hand side of Equation~(\ref{eq:master_moment1_flare}). 
Similarly the expressions give underestimates if the mean 
energy decreases substantially.

We can also deduce general analytic results for the model waiting-time 
and frequency-energy distributions. Substituting the
time-dependent flare-like choice Equation~(\ref{eq:pl_alpha_tdep}) 
into the general expressions for the total flaring rate, 
Equation~(\ref{eq:lambda}), and for the flare frequency-energy 
distribution, Equation~(\ref{eq:ffe}), gives
\begin{equation}\label{eq:lambda_tdep}
\lambda (E,t)=\alpha_0 (t)
    \left(E_c^{-\gamma+1}-E^{-\gamma+1}\right)/(\gamma-1)
\end{equation}
and 
\begin{equation}\label{eq:ffe_tdep}
{\cal N}(E,t)= \alpha_0 (t)
  E^{-\gamma}
  \int_{E}^{\infty}P(E^{\prime},t){\mathrm d}E^{\prime}
\end{equation}
respectively. These expressions apply for $E\geq E_c$: the total rate
and the frequency-energy distribution are both zero if $E<E_c$.

The frequency-energy 
distribution defined by Equation~(\ref{eq:ffe_tdep}) is a power law 
with index $\gamma$ for flare energies less than a time-dependent 
rollover, defined by the largest energy the system is likely to attain 
at a given time. Prior to the step changes considered here, the 
rollover has the approximate lower bound ${\cal E}_1$. After the step 
changes the lower bound is ${\cal E}_{21}$ (for Case~1) and 
${\cal E}_{22}$ (Case~2). If the active region has a very large mean 
energy before and after the change, then most flares involve decreases in 
energy substantially less than the mean energy, and the rollover is not
observed in a short interval of observation. The frequency-energy 
distribution is then given approximately by a simple power law:
\begin{equation}\label{eq:ffe_tdep_bigE}
{\cal N}_{0i}(E)=\alpha_{0i}E^{-\gamma},
\end{equation}
before ($i=1$) and after ($i=2$) the step change. 

The waiting-time distribution given by 
Equation~(\ref{eq:lambda_tdep}) is in general the waiting-time 
distribution for a time-dependent Poisson process with a rate
$\lambda =\lambda [E(t),t]$, which has a well-known form 
\cite{2002SoPh..211..255W}. If the system has a very large mean energy,
so that $E\gg E_c$, Equation~(\ref{eq:lambda_tdep}) may be approximated 
by
\begin{equation}\label{eq:lambda_tdep_bigE}
\lambda (E,t)=\alpha_0 (t)E_c^{-\gamma+1}/(\gamma -1),
\end{equation}
and time dependence enters only through the coefficient $\alpha_0 (t)$.
In this case the total flaring rate has the approximate 
constant values 
\begin{equation}\label{eq:lambda_step_bigE}
\lambda_{01}=\alpha_{01}E_c^{-\gamma+1}/(\gamma -1)
\quad \mbox{and} \quad
\lambda_{02}=\alpha_{02}E_c^{-\gamma+1}/(\gamma -1)
\end{equation}
before and
after the step change respectively, so the waiting-time distribution 
is a simple
exponential before and after the change, corresponding
to Equation~({\ref{eq:wtd_poiss}}) (the waiting-time distribution for
a time-independent Poisson process). Before the step change
the exponent in the exponential is 
$\lambda_{01}$, and after it is $\lambda_{02}$. The 
waiting-time distribution constructed for events both before and 
after the step change is in general a double exponential, specified by 
Equation~(\ref{eq:wtd_pconst-poiss}) with two intervals and rates.
For Case~2, the waiting-time distribution has
the same exponential form (with exponent $\lambda_{01}$)
before and after the step change, because 
the flare transition rate coefficient does not change. If the system
mean energy is small, the distributions may depart from exponential
forms, as discussed in Section~\ref{subsec:ModStats}.

\section{Stochastic modeling}
     \label{sec:Stochastic} 

To illustrate the time-dependent model, and to confirm the qualitative
analytic results given above, we present Monte Carlo solutions to the 
stochastic differential equation formulation 
[Equations~(\ref{eq:stoch_de})--(\ref{eq:hnorm})], following the
approach presented in \inlinecite{2009SoPh..255..211W}. 
We consider only Case~1, since it exhibits the more interesting event 
statistics, and provides a simple model for the observed behavior of
active region AR 11029 (Section~\ref{subsec:TwoARs}).

\subsection{Numerical method}
 \label{subsec:NumMeth} 

Full details of the Monte Carlo method are given in 
\inlinecite{2009SoPh..255..211W}, and here we present only a brief
description appropriate for the step change modeling of 
Case~1. 

A simulation is started with an energy $E_s$ at time $t_s$. The system
is assumed to have a constant flare transition-rate coefficient 
$\alpha_{01}$ and a constant energy-supply rate coefficient $\beta_{01}$.
Prior to the first flare the system evolves in energy according to 
Equation~(\ref{eq:stoch_de}) without the loss term, so the energy
as a function of time is
\begin{equation}\label{eq:Etrajectory}
E^{\ast}(t)=E_s+\beta_{01}(t-t_s).
\end{equation}
The total (expected) flaring rate during this time is 
$\lambda^{\ast}=\lambda [E^{\ast}(t),t]$, where $\lambda (E,t)$ is
given by Equation~(\ref{eq:lambda_tdep}) with 
$\alpha_0 (t)=\alpha_{01}$. A random Poisson waiting time $\Delta t$ is 
generated corresponding to this rate 
\cite{2006SoPh..238...73W,2009SoPh..255..211W}, which defines the end
time $t_e=t_s+\Delta t$ when a jump transition occurs. The 
energy prior to the jump is $E_e=E^{\ast}(t_e)$. A random jump of size 
$\Delta E$ is generated from the distribution $h(\Delta E,E_e,t_e)$ 
defined by Equation~(\ref{eq:h}), as explained in 
\inlinecite{2009SoPh..255..211W}. Once $\Delta E$ is calculated, 
the whole process is repeated, with the new starting time 
$t_s=t_e$ and the new starting energy $E_s=E_e-\Delta E$. The process
is repeated $n_1$ times to give a time history of energy over this
number of jump transitions for the interval prior to the step change.
The time of the last transition, which we label $t_1$, 
defines the time $T$ of the step change. The whole process is then 
repeated again $n_2$ times, with the new flare transition-rate 
coefficient $\alpha_{02}$, to give a time history for the interval 
$t_2$ after the step change. The starting energy for the system 
following the step change is the energy after the last jump transition 
prior to the step change.

This process represents a single simulation. Ensemble averages over
repeated simulations allow comparison with the master equation 
formulation of the model, and the moment equation.
 
For the purposes of the simulations we introduce non-dimensional
parameters
\begin{equation}\label{eq:ndim_param}
\overline{E}=\frac{E}{E_c}, \quad \overline{t}=\frac{t}{t_s}, \quad 
  \overline{\alpha}_{0i}=\frac{\alpha_{0i}t_s}{E_c^{\frac{1}{2}}},
  \quad \overline{\beta}_{01}=\frac{\beta_{01}t_s}{E_c}
\end{equation}
where $t_s$ is an arbitrary scale time, and the choice $\gamma =1.5$
is explicitly shown. We choose $t_s=E_c/\beta_{01}$ without loss
of generality, so that $\overline{\beta }_{01}=1$. The non-dimensional
mean energies for Case~1 are then
\begin{equation}\label{eq:Emean_ndim}
\overline{\cal E}_1=1/\overline{\alpha}_{01}^{2} 
  \quad \mbox{and} \quad
  \overline{\cal E}_{12}=1/\overline{\alpha}_{02}^{2},
\end{equation}
and the estimate for the relaxation time is 
\begin{equation}\label{eq:tau1_ndim}
\overline{\tau}_1 \approx 
  \frac{\overline{\alpha}_{01}+\overline{\alpha}_{02}}
  {4\overline{\alpha }_{01}\overline{\alpha }_{02}^2}.
\end{equation}

\subsection{Results}
 \label{subsec:Results} 

A simulation of Case~1 is performed. The chosen parameters are 
$\overline{\alpha}_{01}=5\times 10^{-3}$ and 
$\overline{\alpha}_{02}=5\times 10^{-2}$  [also 
$\overline{\beta}_{01}=1$, as explained following 
Equation~(\ref{eq:ndim_param})]. These values are chosen to mimic 
solar active region AR 11029 \cite{2010ApJ...710.1324W}, which 
exhibited an approximate ten-fold increase in its flaring rate on 26 
October 2009 (see Section~\ref{subsec:TwoARs} and 
Figure~\ref{fig:dists_ar11029}). The corresponding
approximate mean energies are $\overline{\cal E}_1= 10^{4}$ and
$\overline{\cal E}_2= 10^2$, the approximate total flaring rates [given
by Equation~(\ref{eq:lambda_step_bigE})] are
$\overline{\lambda}_{01}=2\overline{\alpha}_{01}=10^{-2}$ and 
$\overline{\lambda}_{02}=2\overline{\alpha}_{02}=10^{-1}$, and 
the approximate relaxation time is 
$\overline{\tau}_1= 1.1\times 10^{3}$.

\begin{figure}[here]
\centerline{\includegraphics[width=\textwidth]{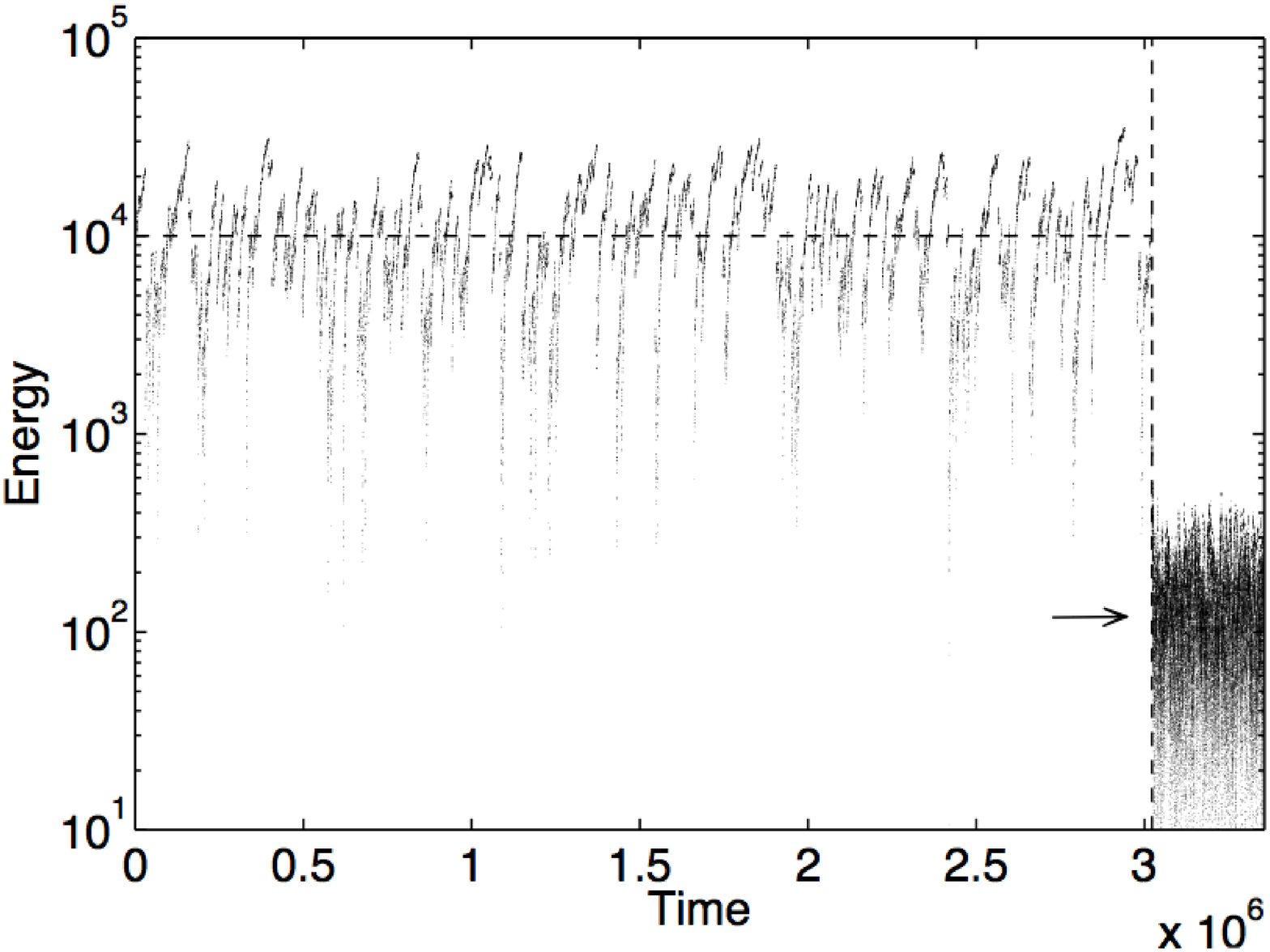}}
\caption{A Monte Carlo simulation of Case~1 showing the system 
energy versus time, for the choice of parameters
$\overline{\alpha}_{01}=5\times 10^{-3}$ and 
$\overline{\alpha}_{02}=5\times 10^{-2}$. A total of $3\times 10^4$ 
events are simulated before and after the step change. The 
dashed horizontal line and the arrow show the mean energies before 
and after the change, respectively, and the time of the change 
is indicated by the dashed vertical line.}
\label{fig:Marko1}
\end{figure}

Figure~\ref{fig:Marko1} shows the results of the simulation, as
a plot of system energy versus time. The mean energies of the system 
before and after the step change are indicated by a dashed
horizontal line and an arrow, respectively, and the time of the 
change is indicated by the dashed vertical line. 
The simulation starts at time $\overline{t}=0$ with the system at 
the initial mean energy estimate $\overline{\cal E}_1= 10^{4}$. 
A total of $n_1=3\times 10^{4}$ flare events are simulated prior 
to the step change with the flare transition-rate coefficient 
$\overline{\alpha}_{01}$, and 
Figure~\ref{fig:Marko1} shows the time history of the energy over 
this number of flares. The final jump transition occurs at time 
$\overline{T}=\overline{t}_1\approx n_1/
  \overline{\lambda}_{01}=3\times 10^6$. 
A total of $n_2=3\times 10^4$ flares
are also simulated after the step change during the interval of time
$\overline{t}_2\approx n_2/
  \overline{\lambda}_{02}=3\times 10^5$, and the time history of energy
is again shown. 
The relaxation time estimate 
$\overline{\tau}_1= 1.1\times 10^{3}$ is small
compared with the scale of the time axis in Figure~\ref{fig:Marko1}. 
Figure~\ref{fig:Marko1} illustrates the fluctuations of the system 
energy around the mean energy before and after the change, and the 
rapid relaxation of the system produced by a large increase in the 
rate.

\begin{figure}[here]
\centerline{\includegraphics[width=\textwidth]{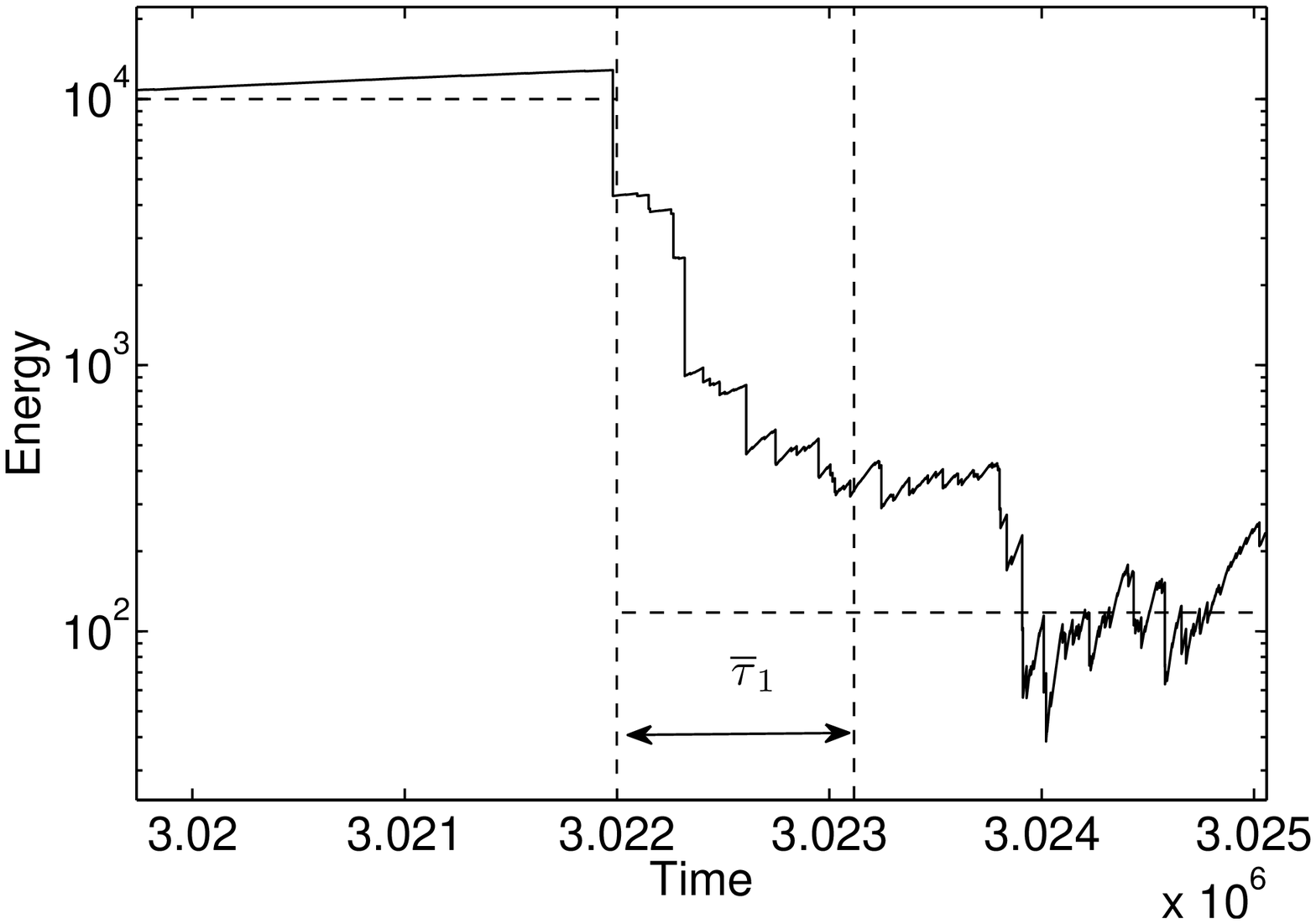}}
\caption{An expanded view in time of the 
evolution of the system energy during the relaxation interval, for the
simulation shown in Figure~\ref{fig:Marko1}. The estimate 
$\overline{\tau}_1= 1.1\times 10^{3}$ for the relaxation time is 
indicated.}
\label{fig:Marko2}
\end{figure}

Figure~\ref{fig:Marko2} shows an expanded view in time of the 
evolution of the system energy during the interval of relaxation, for 
the simulation shown in Figure~\ref{fig:Marko1}. The mean energies
before and after the step change are shown by dashed 
horizontal lines, the time of the step change is shown by the 
dashed vertical line on the left, and the estimate 
$\overline{\tau}_1= 1.1\times 10^{3}$ for the relaxation time is 
indicated by the interval between the dashed vertical lines. This 
figure shows the specific path $E=E(t)$ to a new steady state taken 
in this simulation.

\begin{figure}[here]
\centerline{\includegraphics[width=\textwidth]{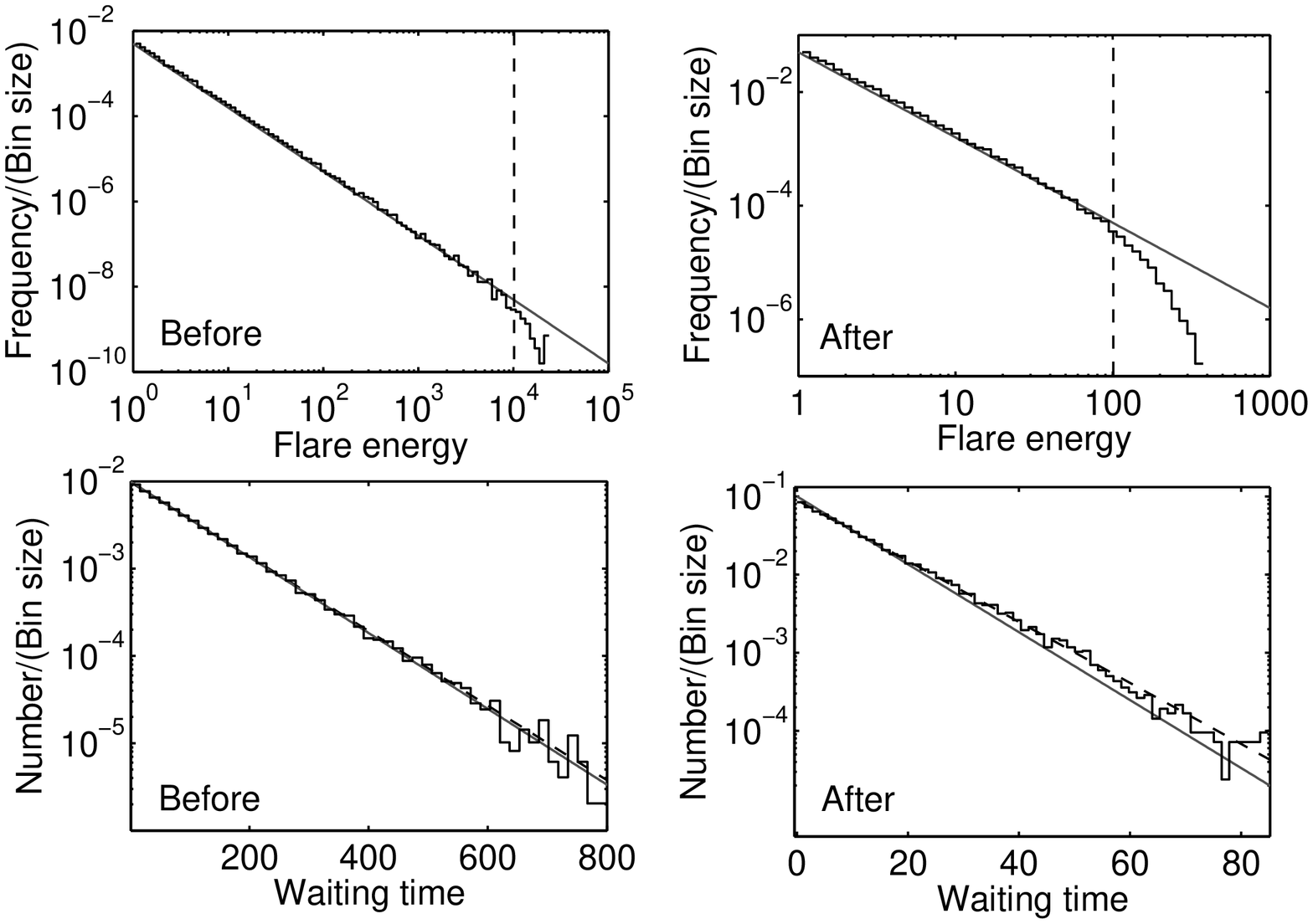}}
\caption{Histograms representing the frequency-energy (upper row), 
and waiting-time (lower row) distributions before (left column)
and after (right column) the step 
change, for the simulation shown in Figures~\ref{fig:Marko1} 
and~\ref{fig:Marko2}. The dashed vertical lines in the upper row
are the mean energy estimates and the solid grey curves are the
power-law model for an active region with a a very large mean energy. 
The solid grey lines in the lower row are the 
the simple exponential (Poisson) models corresponding to the rates
$\overline{\lambda}_{01}$ and $\overline{\lambda}_{02}$, and the 
dashed lines are the exponential models with rates 
$\overline{\lambda}_{s1}$ and $\overline{\lambda}_{s2}$ (see text).}
\label{fig:Marko3}
\end{figure}

Figure~\ref{fig:Marko3} presents the histograms of the 
frequency-energy distributions (upper row) and the waiting-time
distributions (lower row) for the simulation shown in 
Figures~\ref{fig:Marko1} and~\ref{fig:Marko2} before (left column) 
and after (right column) the step change. The frequency-energy
distributions are plotted in a log-log representation, with the mean 
energy estimates 
$\overline{\cal E}_1= 10^{4}$ and $\overline{\cal E}_2= 10^2$ shown 
by the dashed vertical lines, and the simple power-law models for a 
system with a very large free energy [given by 
Equation~(\ref{eq:ffe_tdep_bigE})] shown by the solid grey lines. 
The frequency-energy distributions follow the expected
power laws for energies less than a rollover energy which is 
comparable to the mean energy estimate in each case. These results
confirm the analytic arguments given in 
Section~\ref{subsec:AnalyticTDep}. The waiting-time histograms
in the lower row of Figure~\ref{fig:Marko3} are shown together with 
the simple exponential distributions corresponding to the estimates
$\overline{\lambda}_{01}=0.01$ and $\overline{\lambda}_{02}=0.1$ for the
mean rates of events (solid grey lines).
The results show that the waiting times follow approximate
exponential (Poisson) distributions, as deduced by the analytic 
arguments in Section~\ref{subsec:AnalyticTDep}. However, there is
a significant discrepancy between the slope of the simulation 
histogram and the slope of the simple model (corresponding to the
rate estimate $\overline{\lambda}_{02}=0.1$) after the change. 
In the simulation,
the mean rates defined by the number of events divided by the elapsed
time are $\overline{\lambda}_{s1}=n_1/\overline{t}_1=0.0099$ (before 
the change) and $\overline{\lambda}_{s2}=n_2/\overline{t}_2=0.090$
(after). The rate after the change is significantly less than
the estimate given by
Equation~(\ref{eq:lambda_step_bigE}), which indicates that the 
approximation $E\gg E_c$ is not being met. Before the change, 
the mean energy is approximately ${\cal E}_1=10^4E_c$, and 
Equation~(\ref{eq:lambda_step_bigE}) provides a good approximation.
After the change ${\cal E}_1=10^2E_c$, and the approximation is
poorer. The rate is reduced because the low system energy prevents 
larger flares from occurring.
However, the functional form of the waiting-time distribution
remains approximately exponential. The lower row in 
Figure~\ref{fig:Marko3} also shows the simple exponential models 
corresponding to the mean rates $\overline{\lambda}_{s1}$ and 
$\overline{\lambda}_{s2}$ (dashed lines), and there is good agreement 
in both cases with the simulation histograms.

\begin{figure}[here]
\centerline{\includegraphics[width=\textwidth]{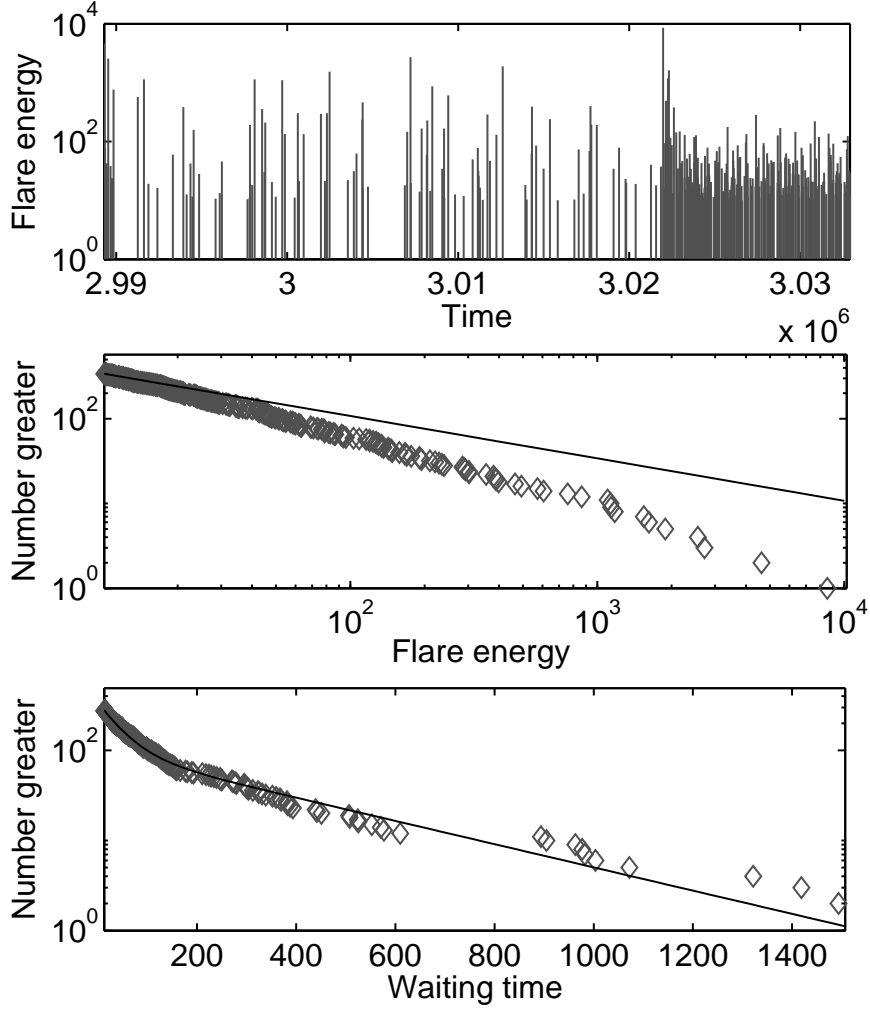}}
\caption{Events in the simulation shown in 
Figures~\ref{fig:Marko1}-\ref{fig:Marko3} larger than size
$\overline{E}_1=10$ which occur in the interval between 30 relaxation 
times before, and ten relaxation times after, the change in the rate. 
The format of the figure is the same as for the observational data
for active region AR 11029 shown in Figure~\ref{fig:dists_ar11029}.}
\label{fig:Marko4}
\end{figure}

Figure~\ref{fig:Marko4} illustrates part of the data from the same
simulation, and is intended for qualitative comparison with the 
observational data for active region AR 11029, shown in 
Figure~\ref{fig:dists_ar11029}. The format of the figure is the same
as Figure~\ref{fig:dists_ar11029}. The figure shows the events from
the simulation with energy larger than $\overline{E}_1=10$ which 
occurred in the interval of time between 
$\overline{t}=\overline{T}-30\overline{\tau}_1$ and
$\overline{t}=\overline{T}+10\overline{\tau}_1$, where 
$\overline{T}=3.022\times 10^6$ is the time of the change and
$\overline{\tau}_1= 1.1\times 10^{3}$ is the relaxation-time estimate.
These choices are intended to crudely mimic observational event 
selection above a threshold, and to include numbers of events before
and after the change which provide good statistics but which are 
comparable with observational numbers for the soft X-ray events. 
The choices give $N_1=341$ events in
total, with $n_{11}=98$ events before the change and $n_{12}=243$ 
after.
The upper panel in Figure~\ref{fig:Marko4} shows the flare energies 
versus time, in the format of Figure~\ref{fig:dists_ar11029}. The 
middle panel shows the the flare cumulative number distribution for 
the selected events (grey diamonds), which corresponds to the 
frequency-energy distribution according to Equation~(\ref{eq:cum_size}). 
The black line is the simple power-law model for a system with a 
very large free energy, corresponding to 
Equation~(\ref{eq:ffe_tdep_bigE}). Specifically, the black line is
$\overline{C}_0(\overline{E})=
  N_1(\overline{E}/\overline{E}_1)^{-\gamma+1}$.
The simulation events show a departure from the model 
$\overline{C}_0(\overline{E}_1)$ at large energy due to the influence of the 
rollovers $\overline{\cal E}_1$ and $\overline{\cal E}_2$ (see the 
upper row of Figure~\ref{fig:Marko3}). The observed rollover for the
selected events is qualitatively similar to the observations for 
active region AR 11029 (middle panel of 
Figure~\ref{fig:dists_ar11029}). The lower panel in 
Figure~\ref{fig:Marko4} shows the cumulative waiting-time distribution
for the selected events (grey diamonds), and the double exponential
Poisson model (black curve) given by 
Equation~(\ref{eq:wtd_pconst-poiss}):
\begin{equation}\label{eq:sim_two_rate_mod_wtd}
\overline{P}(\Delta \overline{t})=
\frac{n_{11}}{N_1}\overline{\lambda}_{11}
\exp (-\overline{\lambda}_{11}\Delta\overline{t})
+\frac{n_{12}}{N_1}\overline{\lambda}_{12}
\exp (-\overline{\lambda}_{12}\Delta\overline{t}),
\end{equation}
where 
\begin{equation}
\overline{\lambda}_{11}=\frac{n_{11}}{30\overline{\tau}_1}
\quad \mbox{and} \quad
\overline{\lambda}_{12}=\frac{n_{12}}{10\overline{\tau}_1}
\end{equation}
are the rates of selected events before and after the change. This 
panel, which may be compared qualitatively with the lower panel in 
Figure~\ref{fig:dists_ar11029}, shows that the selected events follow
the double power-law model.

\begin{figure}[here]
\centerline{\includegraphics[width=\textwidth]{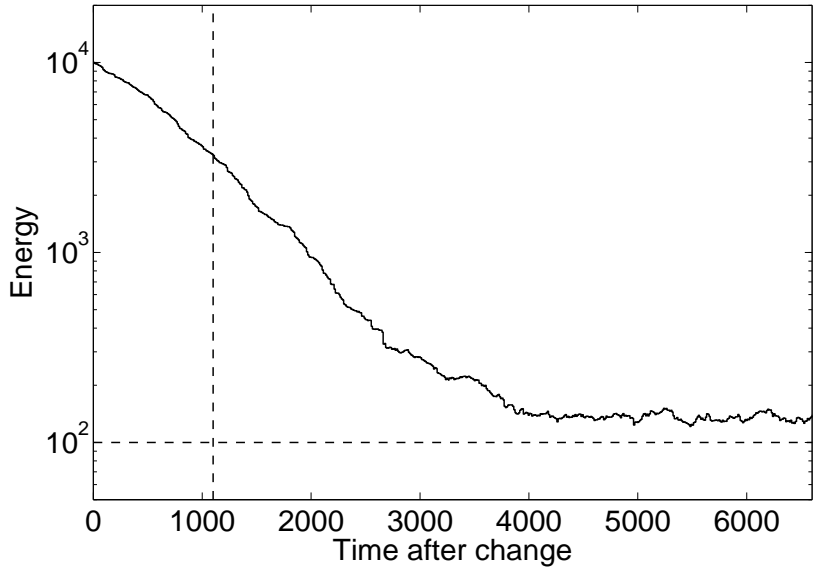}}
\caption{An average of the energy of the system versus time
over an ensemble of 100 simulations constructed for the time after 
the step change. The simulations start at the mean energy estimate
$\overline{\cal E}_1=10^4$ for the time before the 
change. The estimate for the mean energy 
$\overline{\cal E}_2=10^2$ after the change is shown by the solid 
horizontal line, and the dashed vertical line is the estimate
$\overline{\tau}_1= 1.1\times 10^{3}$ for the relaxation time.}
\label{fig:Marko5}
\end{figure}

Figure~\ref{fig:Marko5} illustrates the relaxation process using an 
ensemble of 100 simulations each of $n_2=3\times 10^4$ events generated
for the time after the step change. 
In each simulation the system is started at time $\overline{t}=0$
with energy equal to the estimate $\overline{\cal E}_1=10^4$ 
for the mean energy of the system {\em before} the change, and flare 
events are simulated for the interval of time 
$0\leq \overline{t}\leq 6\overline{\tau}_1$. These choices
simulate the relaxation of the system from the steady state 
existing before the change in the rate.
The average energy over the ensemble of 100 simulations is calculated
for 600 equally spaced times during the simulation interval. 
Figure~\ref{fig:Marko5} shows the ensemble-mean energies versus time.
The estimate $\overline{\tau}_1= 1.1\times 10^{3}$ for 
the relaxation time is indicated by the dashed vertical line. The
average energy decreases towards the mean energy estimate
$\overline{\cal E}_2=10^2$ (dashed horizontal line) following the 
change, reaching a statistically steady state after a time 
$\overline{\tau}_{\rm sim}\approx 4000-5000\approx 4\overline{\tau}_1$.
These results suggest that the estimate $\overline{\tau}_1$
underestimates the actual relaxation time, which is not surprising 
given the large increase in the flaring rate, and hence decrease in 
the mean energy of the system [see comments following 
Equation~(\ref{eq:tau2})]. This figure also indicates that
the estimate for the final mean energy is an underestimate. This 
is also not surprising as it is an approximation, and in particular
involves the assumption ${\cal E}_2\gg E_c$ [see 
\inlinecite{2008ApJ...679.1621W} for the derivation].

\section{Conclusions}
     \label{sec:Conclusions} 

A stochastic description of the free energy of an active region,
a `jump-transition model'
\cite{1998ApJ...494..858W,2008ApJ...679.1621W,2009SoPh..255..211W} 
is applied to model an active region which exhibits time variation 
in its mean flaring rate. Time variation in flare productivity is 
commonly observed on the Sun, for example in NOAA solar active region 
AR 11029, which on October 26 exhibited an approximate 10-fold 
increase in flaring, based on soft X-ray events compiled by the US 
Space Weather Prediction Center \cite{2010ApJ...710.1324W}. The 
general jump-transition model has previously been investigated only for 
time-independent situations, in which case it reproduces observed
solar flare statistics (in particular the power-law flare frequency 
energy distribution, and a Poisson waiting-time distribution). The 
interest here is with whether the time-dependent generalization of the 
model also succeeds in this respect.

Time variation is incorporated in the model for two simple cases:
1.\ a step change in a coefficient $\alpha_0 (t)$ describing the 
rates of flaring; and 2.\ a step change in a coefficient $\beta_0(t)$
describing the mean rate of energy of the system. Case 1 may be 
appropriate to describe solar active region AR 11029. Analytic 
arguments are presented that predict the qualitative behavior of the 
model in response to the changes. In both cases the system adjusts by 
shifting to a new
statistically stationary steady state following the change, over a 
relaxation time, which is analytically estimated. Steady states of 
the system have mean energies which may also be analytically 
estimated, as shown in \inlinecite{1998ApJ...494..858W}. Provided the
mean energy is sufficiently large, the steady-state system 
exhibits a power-law flare frequency energy distribution (for flare 
energies less than a rollover value corresponding approximately to 
the mean energy),
and exhibits simple Poisson waiting-time statistics, as shown in 
\inlinecite{2008ApJ...679.1621W} and \inlinecite{2009SoPh..255..211W}. 
The analytic arguments given here predict that the same is true in
time-dependent situations, although the waiting-time distribution
corresponds to a time-dependent Poisson process with a rate
determined by the model coefficient $\alpha_0 (t)$. For Case 1, the
waiting-time distribution is in general a double exponential, 
corresponding to two intervals of flaring with a constant rate.

Monte Carlo simulations of the stochastic differential equation
formulation of the jump transition model \cite{2009SoPh..255..211W} 
are presented for Case 1, for parameter choices intended to
qualitatively model the soft X-ray event data for active region 
AR 11029. The simulations confirm the analytic arguments for the 
behavior of the model. The observed soft X-ray flare frequency-peak 
flux distribution for this active region showed evidence for an upper
rollover \cite{2010ApJ...710.1324W}, which was interpreted in terms 
of the finite amount of energy 
available for flaring in a small active region being depleted by an 
interval of very rapid flare production. This interpretation is 
consistent with the the jump-transition model, and the rollover is
qualitatively reproduced in the simulations presented here. 
A double exponential waiting-time distribution is obtained in the 
simulations which mimics that observed for solar active region 
AR 11029. 

A quantitative comparison of data such as that for active region AR 
11029 with the jump transition model requires estimation of flare 
energies, and active region energies, from the data. However, soft 
X-ray peak fluxes are indicative only of flare energy, and we are 
currently unable to reliably estimate the magnetic free energy of 
a solar active region [see e.g.\ \inlinecite{2009ApJ...696.1780D} 
for an attempt to obtain energy estimates for one active region via 
magnetic field modeling]. Nevertheless, in future work we will 
consider more detailed comparison of the model with observational
data.

An interesting aspect of the jump-transition model in this context is
the model prediction of a departure from Poisson waiting-time 
statistics, if the system energy becomes sufficiently small, 
because large events are prevented from occurring, which reduces the 
overall mean rate of events. This can lead to departure from an
exponential waiting-time distribution, even in the time-independent
case \cite{2008ApJ...679.1621W,2009SoPh..255..211W}. However, the 
simulations presented here reveal a new effect. If the system mean 
energy has intermediate values, the overall rate of events is reduced 
by comparison with that for a system with a large mean energy, but the
waiting-time distribution remains approximately exponential. Hence
the waiting-time distribution may remain exponential even when there
is detailed departure from Poisson statistics.
This effect was not noticed in previous
modeling \cite{2008ApJ...679.1621W,2009SoPh..255..211W}, and also 
warrants further investigation. The soft X-ray events in AR 11029 
appeared to follow Poisson occurrence in time based on agreement 
between the waiting-time distribution and the Poisson model 
\cite{2010ApJ...710.1324W}. However, the US SWPC 
event data suffer from significant event detection and selection 
problems due to the large and time varying soft X-ray background 
[discussed, for example, in \inlinecite{2001SoPh..203...87W}]. 
Comparison with more sensitive data may be required to reveal detailed
departure from Poisson occurrence. 

The advent of new high resolution and high cadence short-wavelength
observations of the Sun by the Advanced Imaging Assembly (AIA) on 
the Solar Dynamics Observatory (SDO) should enable improved 
investigations of solar flare event statistics, and more careful 
comparison with flare statistics models such as the one
in this paper. The results may provide new insights into the mechanisms
of energy storage and release underlying the flare phenomenon.


\begin{acks}
The authors thank Kinwah Wu for comments on a draft of the paper.
\end{acks}


\end{article} 
\end{document}